\documentclass{PoS}

\title{Gluonic Excitations and Experimental Hall-D at Jefferson Lab}

\ShortTitle{Gluonic Excitations and Experimental Hall-D at JLab}

\author{\speaker{Justin STEVENS, for the GlueX Collaboration}\\
       Massachusetts Institute of Technology \\
       E-mail: \email{jrsteven@mit.edu}}

\abstract{A new tagged photon beam facility is being constructed in experimental Hall-D at Jefferson Lab as a part of the 12 GeV upgrade program.  The 9 GeV linearly-polarized photon beam will be produced via coherent Bremsstrahlung using the CEBAF electron beam, incident on a diamond radiator.  The GlueX experiment in Hall-D will use this photon beam to search for and study the pattern of gluonic excitations in the meson spectrum produced through photoproduction reactions with a liquid hydrogen target.
Recent lattice QCD calculations predict a rich spectrum of hybrid mesons, that are formed by exciting the gluonic field that couples the quarks.  A subset of these hybrid mesons are predicted to have exotic quantum numbers which cannot be formed from a simple $q\bar{q}$ pair, and thus provide an ideal laboratory for testing QCD in the confinement regime.  In these proceedings the status of the construction and installation of the GlueX detector will be presented, in addition to simulation results for some reactions of interest in hybrid meson searches.}

\FullConference{XXII. International Workshop on Deep-Inelastic Scattering and Related Subjects \\
		 28 April - 2 May 2014 \\
		 Warsaw, Poland}

\begin{document}

\section{Introduction}
The constituent quark model successfully describes the vast majority of observed mesons, grouping them in nonets according to their quantum numbers.  However, only a certain set of quantum numbers - $J^{PC}$: total angular momentum ($J$), parity ($P$) and charge conjugation ($C$) - are allowed for these bound states of a quark ($q$) and antiquark ($\bar{q}$).  Recent lattice QCD calculations~\cite{Dudek:2013yja} predict a rich spectrum of mesons, including a class of states known as hybrid mesons which can be naively though of as a $q\bar{q}$ pair coupled to an excited gluon ($q\bar{q}g$).  A subset of these hybrids, the so-called ``exotic" hybrid mesons, has an unmistakable experimental signature of a $J^{PC}$ that cannot be created from a simple $q\bar{q}$ pair (\textit{e.g.} $0^{+-}, 1^{-+}, 2^{+-}$, etc.).  
A recent review of the experimental measurements of hybrid meson production can be found in Ref.~\cite{Meyer:2010ku}, however there is some contention whether any hybrid states have been unambiguously identified.  The primary goal of the GlueX experiment, as described in this proceedings, is to search for and study these hybrid mesons using the photoproduction process in Hall~D at Jefferson Laboratory (JLab).

\section{GlueX Experiment} 
\label{sec:detector}

\begin{figure} [h]
\begin{center}
\includegraphics[width=0.70\linewidth]{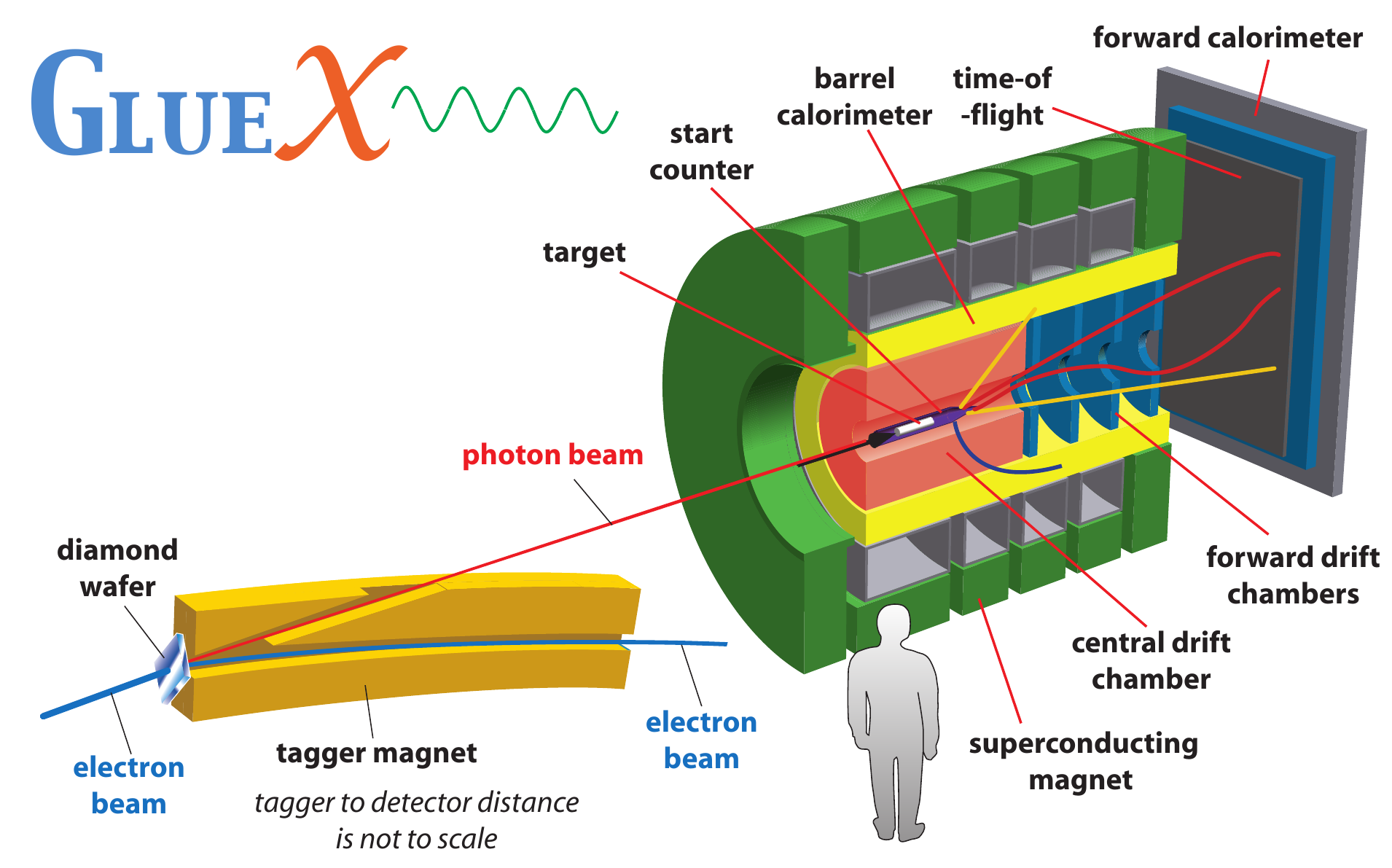}
\caption{\label{fig:detector} A schematic of the GlueX detector and beamline.}
\end{center}
\end{figure}

As a part of the 12 GeV Upgrade program at JLab a new experimental hall, Hall~D, was constructed to house the GlueX experiment, which is schematically shown in Fig.~\ref{fig:detector}.  The experiment utilizes a photon beam derived from the 12 GeV electron beam's coherent bremsstrahlung radiation from a thin (20 ${\mu}m$) diamond wafer.  The coherent bremsstrahlung process provides a linearly-polarized photon beam with a ``coherent peak'' in the photon beam energy that can be tuned by adjusting the diamond lattice angle.  For the GlueX experiment, this coherent peak is between $8.4 < E_\gamma < 9$~GeV, where the linear polarization is expected to be 40\%.  The electrons scattered in the coherent bremsstrahlung process are bent out of the electron beamline by a 1.5~T tagger magnet, where an array of scintillator detectors are used to determine the photon energy with a resolution of 0.1\% in the energy range of the coherent peak.  An early commissioning electron beam was delivered to the Hall~D tagger on May~7, 2014, which represents a significant milestone for the 12 GeV upgrade project at JLab and the GlueX experiment.

The GlueX detector itself consists of a 2~T superconducting solenoid with both central and forward tracking chambers and calorimeters surrounding a liquid hydrogen target~\cite{AlekSejevs:2013mkl}.  The large acceptance of the detector is required to have a high efficiency for many-particle final states with both charged and neutral decay particles.  With the addition of precision charged particle identification, through timing information, this allows one to study many possible decay topologies of the candidate hybrid mesons and provide additional insight into their quark flavor content.

\begin{figure} [t]
\begin{center}
\includegraphics[width=0.45\linewidth]{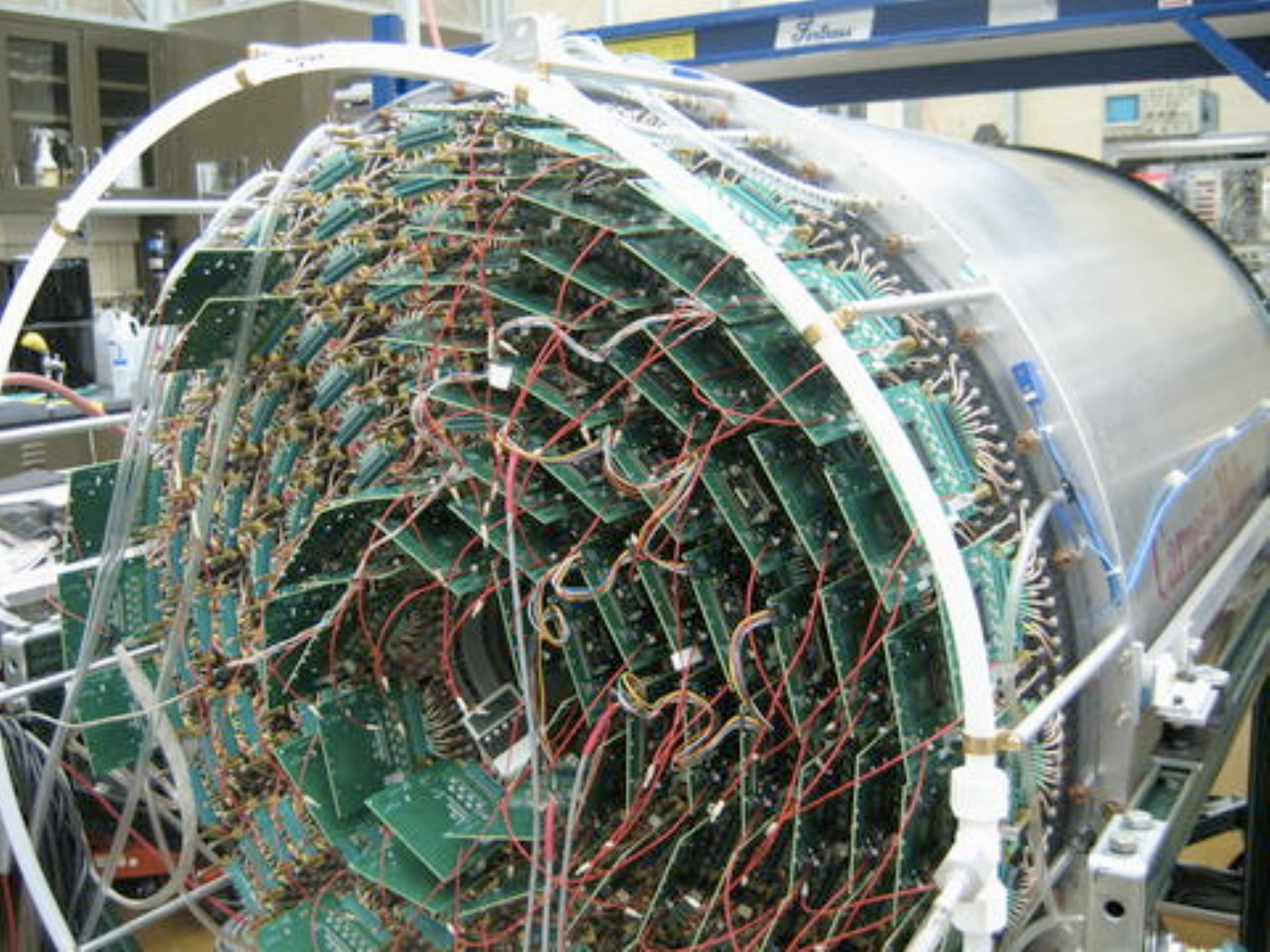}
\includegraphics[width=0.45\linewidth]{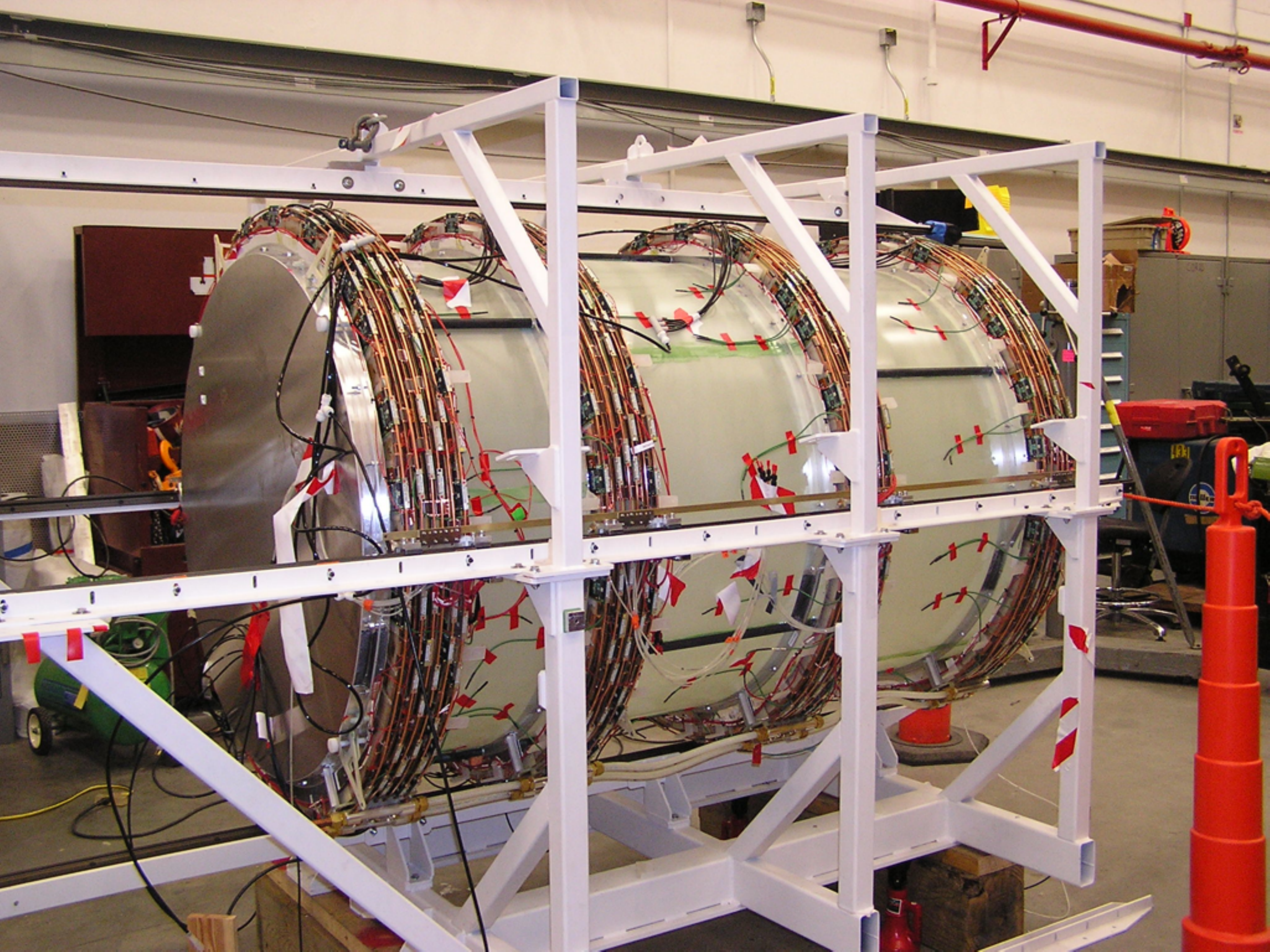}
\caption{\label{fig:tracking} Charged particle tracking in the GlueX experiment is provided by the Central Drift Chamber (left) and Forward Drift Chamber (right), both shown fully constructed before they were installed in the solenoid.}
\end{center}
\end{figure}

Charged particle tracking in GlueX is provided by the Central Drift Chamber (CDC) and Forward Drift Chamber (FDC), which together cover the polar angle range, $1^{\circ} < \theta < 140^{\circ}$.  The CDC and FDC, shown fully constructed in Fig.~\ref{fig:tracking}, will provide position resolutions of roughly 150~${\mu}m$ and 200~${\mu}m$, respectively, resulting in a momentum resolution of 1\% - 3\% for tracks with $\theta > 5^{\circ}$.  Both the CDC and FDC have been constructed and installed, and are currently undergoing early commissioning in preparation for first photon beam running in Fall 2014.

\begin{figure} 
\begin{center}
\includegraphics[width=0.45\linewidth]{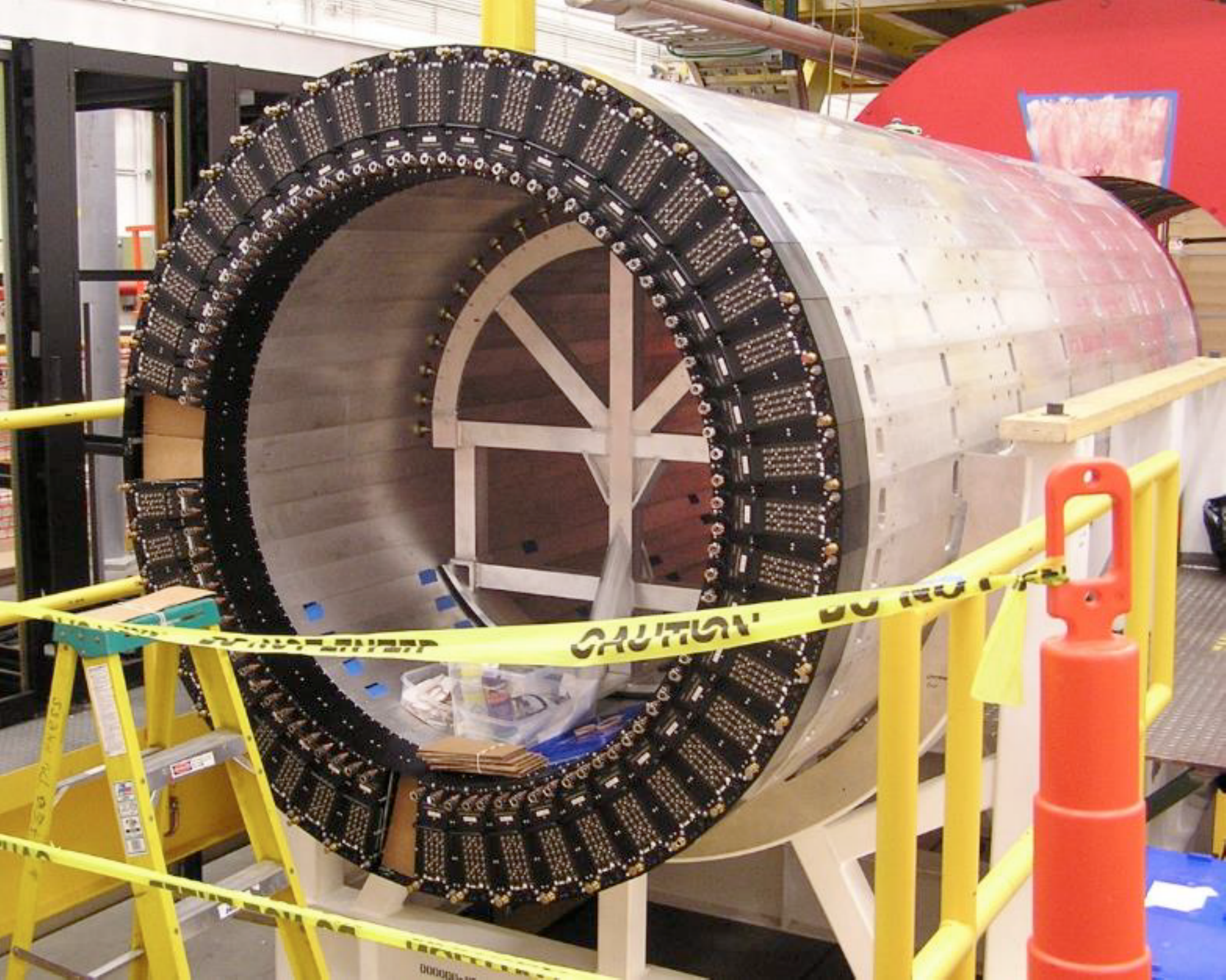}
\includegraphics[width=0.24\linewidth]{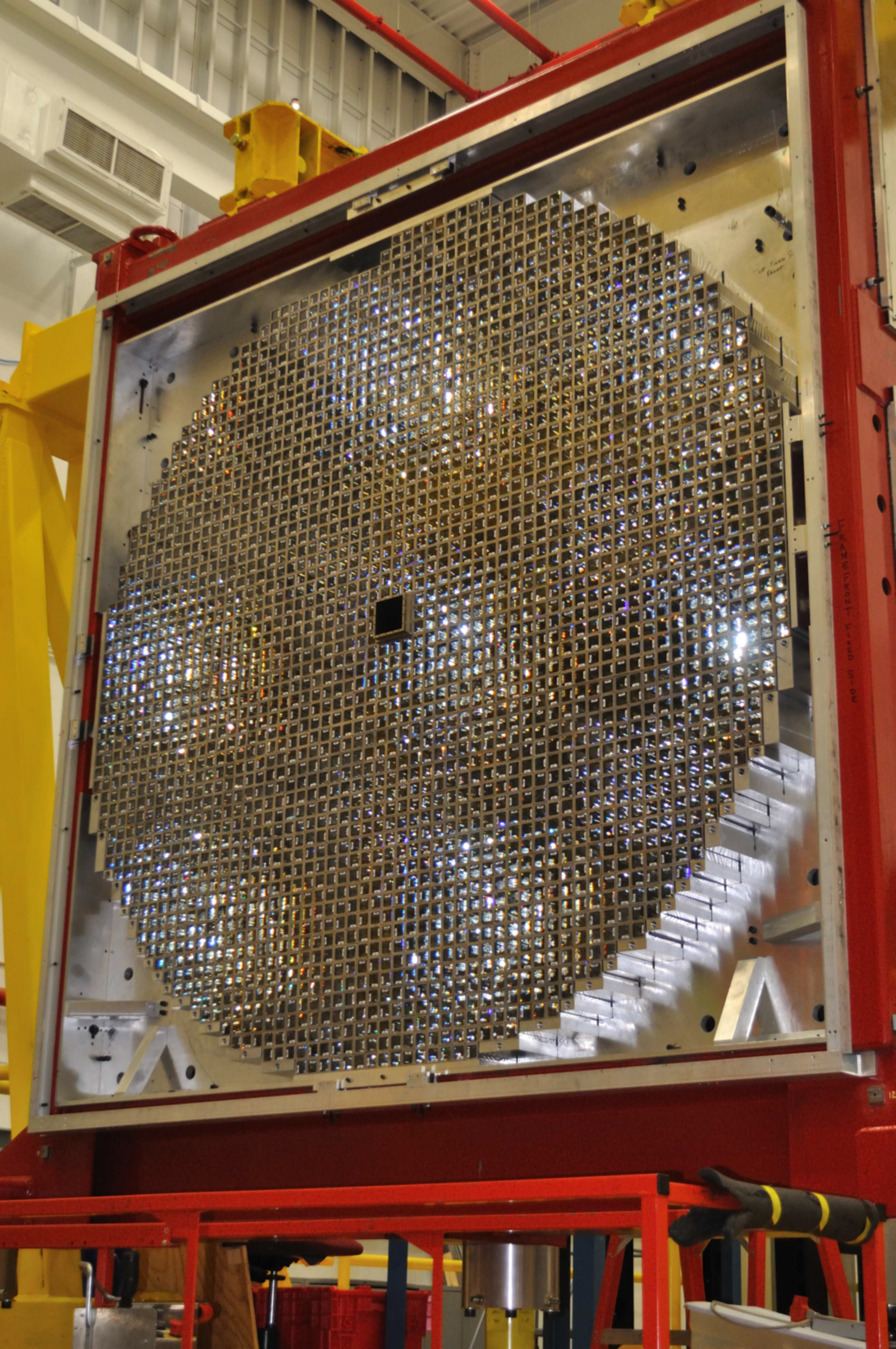}
\caption{\label{fig:calorimetry} (left) The Barrel Calorimeter fully constructed, before being installed inside the solenoid. (right) The Forward Calorimeter.}
\end{center}
\end{figure}

Similar to the drift chambers, the GlueX calorimeters are separated into different angular regions with the Barrel Calorimeter (BCAL) covering $11^{\circ} < \theta < 120^{\circ}$ and the Forward Calorimeter (FCAL) covering $2^{\circ} < \theta < 11^{\circ}$.  The BCAL and FCAL, shown in Fig.~\ref{fig:calorimetry}, are both fully constructed and installed in Hall~D, and are currently being commissioned, including first observations of cosmic rays.  The energy resolutions of the BCAL and FCAL are $5.5\%/\sqrt{E} \pm 2.0\%$ and $5.6\%/\sqrt{E} \pm 3.5\%$, respectively.

Charged particle identification is a critical element of the GlueX experiment, with key information provided by several sub-detectors.  Precision timing is provided in the forward direction by the Time of Flight scintillator wall covering $2^{\circ} < \theta < 11^{\circ}$, with 70~ps timing resolution which results greater than 4$\sigma$ $\pi/K$ separation up to momenta of  ${\sim}$2~GeV/$c$.  Additionally, the Start Counter (consisting of 30 thin scintillator paddles surrounding the target) provides an event start time with a design resolution of 350 ps.  The BCAL provides timing information in the central portion of the detector, and the drift chambers provide complementary energy loss measurements useful for lower momentum charged particles.  

\section{Simulation Studies} 

The expected performance of the GlueX detector to reconstruct and identify particular final state topologies has been evaluated through several studies of large simulation samples of inclusive photoproduction in PYTHIA ($1{\times}10^9$ events).  In addition to the charged and neutral particle reconstruction algorithms, a procedure (known as a ``kinematic fit") has been developed to utilize the over-constrained kinematics of the interaction to reject incorrectly reconstructed background events.  This kinematic fit, along with particle identification and other information have been utilized in Boosted Decision Tree (BDT)~\cite{BDT} analyses to select events with particular final state topologies.  A BDT is a multivariate classifier which utilizes machine learning with samples of known signal or background simulation events to ``train" a classifier, which can be used to select a particular final state topology from experimental data.  In a study of the reaction $\gamma p \rightarrow n \pi^+ \pi^- \pi^+$, a selection efficiency of 22\% was achieved with a purity of 95\%.  

Studies of the kaon identification capabilities of the GlueX detector have been carried out as a part of a recent JLab PAC proposal~\cite{AlekSejevs:2013mkl}.  The signal efficiencies expected for different purity levels are given in Table~\ref{tab:bdt_eff} for the baseline GlueX detector, described in Sec.~\ref{sec:detector}.  An upgrade to the particle identification capabilities of the GlueX detector has been proposed to utilize a portion of the BaBar DIRC detector to provide enhanced $\pi/K$ separation at higher momentum.  The expected performance improvements from this focusing DIRC (FDIRC) detector are given in Table~\ref{tab:bdt_eff}.

\begin{table*}
\centering
\begin{tabular}{c|cc|cc|cc|cc} \hline\hline
~~~~~~~~~~~~  & \multicolumn{2}{c}{$\eta_{1}^{\prime}(2300)\to K^*K_S$} & \multicolumn{2}{c}{$h_{2}^{\prime}(2600)\to K_1^+K^-$} & \multicolumn{2}{c}{$\phi_{3}(1850)\to K^+K^-$} &  \multicolumn{2}{c}{$Y(2175)\to \phi f_0(980)$} \\ \hline
Purity & Baseline & FDIRC & Baseline & FDIRC &  Baseline & FDIRC & Baseline & FDIRC\\
~0.90~ & ~0.36 & ~0.48 & ~0.33 & ~0.49 & ~0.67 & ~0.74 & ~0.46 & ~0.65 \\
~0.95~ & ~0.18 & ~0.33 & ~0.16 & ~0.34 & ~0.61 & ~0.68 & ~0.20 & ~0.55 \\ 
~0.99~ & ~0.00 & ~0.05 & ~0.00 & ~0.08 & ~0.18 & ~0.38 & ~0.03 & ~0.28 \\
\hline\hline
\end{tabular}
\caption[]{\label{tab:bdt_eff} Efficiencies for identifying several hybrid meson candidate final states in GlueX excluding reconstruction of the final state tracks.}
\end{table*}

\begin{figure}
\begin{center}
\includegraphics[width=1.0\linewidth]{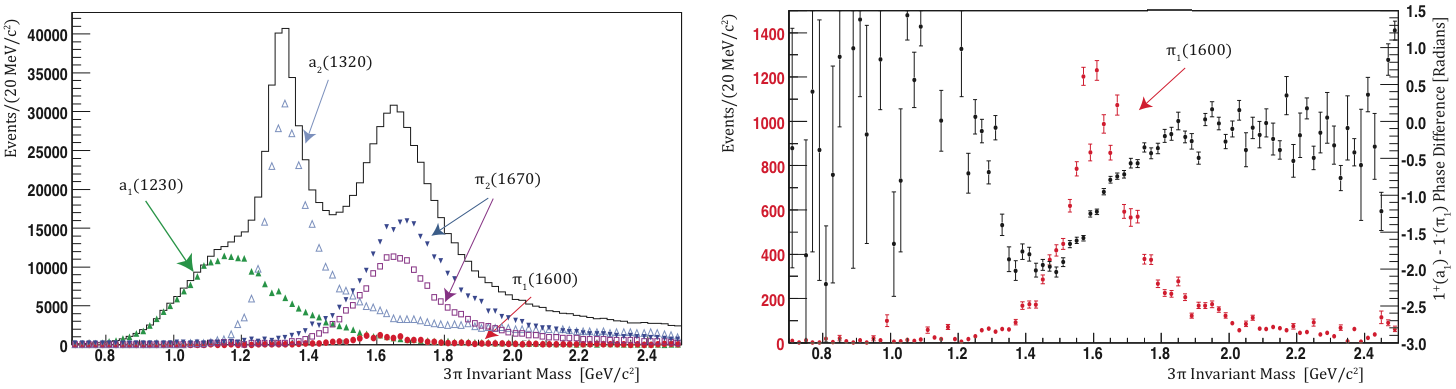}
\caption{\label{fig:amp_analysis} (left) The invariant mass spectrum as a function of $M(\pi^+\pi^-\pi^+)$ is shown by the solid histogram.  The results of the amplitude decomposition into resonant components in each bin is shown with points and error bars.  (right) The exotic $\pi_1(1600)$ amplitude is cleanly extracted (red points).  The black points show the $\pi_1$ phase relative to a fixed phase amplitude.}
\end{center}
\end{figure}

In addition to selecting pure samples of particular final state particle topologies, an amplitude analysis must be performed to extract the $J^{PC}$ quantum numbers of the observed resonances.  Figure~\ref{fig:amp_analysis} shows an example of such an analysis using simulated data from the GlueX experiment for the reaction $\gamma p \rightarrow n \pi^+ \pi^- \pi^+$~\cite{AlekSejevs:2013mkl}.  In the $\pi^+\pi^-\pi^+$ mass distribution, Fig.~\ref{fig:amp_analysis} (left), several conventional resonances were generated with an additional exotic hybrid, $\pi_1(1600)$ in red, included with a relative strength of 1.6\%.  The exotic amplitude and its phase are clearly extracted with data of this statistical precision, which corresponds to only 3.5 hours of high intensity GlueX running.

\section{Conclusion and Outlook}
In summary, the GlueX experiment in Hall D at Jefferson Laboratory is designed to search for and study the spectrum of light hybrid mesons, for which a rich spectrum is predicted by Lattice QCD.  Construction and installation of the detector components is nearing completion, and the first photon beam for GlueX detector commissioning is expected in Fall 2014, with initial physics running planned for 2015.  In addition to the approved beam time for GlueX in Hall~D, there are two other approved proposals, the PrimeX~\cite{PACPrimeX} and Charged Pion Polarizablility~\cite{PACCPP} experiments, which are designed to perform specific precision tests in low energy QCD.

\end{document}